\documentclass[twoside, epsfig]{article}
\input oejv.sty

\begin{document}
\OEJVhead{Month year}
\OEJVtitle{Search for period changes in Mira stars}
\OEJVauth{Nesci, Roberto$^1$ and Rocchi, Gianni$^2$ }
\OEJVinst{INAF/IAPS-Roma, via Fosso del Cavaliere 100, 00133 Roma, Italy, 
{\tt \href{mailto:roberto.nesci@inaf.it}{roberto.nesci@inaf.it}}}
\OEJVinst{Gruppo Astrofili Monte Subasio, via Don Luigi Barabani 31, 06018 Porziano (PG), Italy}

\OEJVabstract{We reobserved in the $R_C$ and $i'_{Sloan}$ bands, during the years 2020-2021, seven Mira variables in Cassiopeia, for which historical $i'_{Sloan}$ light curves were available from Asiago Observatory plates taken in the years 1967-84. The aim was to check if any of them had undergone a substantial change in the period or in the light curve shape. Very recent public data form ZTF-DR5 were also used to expand our time 
base window. A marked color change was detected for all the stars along their variability cycle. The star V890 Cas showed a significant period  decrease of 12\% from 483 to 428 days, one of the largest known to date. All the stars, save AV Cas, showed a smaller variation amplitude in the recent CCD data, possibly due to a photometric accuracy higher than that of the photographic plates.
}
\begintext
\section{Introduction}\label{secintro} 
In the years 1967-75, at the Asiago Observatory, a program of search of Mira variables was carried on using infrared plates (hypersensitized Kodak I-N emulsion + RG5 filter) and the Schmidt 60/90/215 telescope centered on Gamma Cas. A few further plates were taken in the years 1981-84. However, only recently \citep{Nesci18} these plates were actually studied using an automatic pipeline, and 21 long period variable stars were found, 9 of them being new discoveries. Spectral types were also obtained for all of them.
For the 10 Mira stars in the sample, a robust period determination was obtained, given the large time interval covered. Despite many of them were already formally listed in the VSX or GCVS catalogs, only 4 had a published period and none was studied in detail before. All these stars have a steep spectral shape peaking in the near infrared, so are not easily observed by amateur astronomers or by robotic surveys like ASAS-SN \citep{Schap14}, operating in the blue and visual bands. 

It is well known, since the extensive search by \citet{Templeton05} of period variation in a sample of 540 Miras, observed for about a century by the AAVSO observers, that only very few Miras (8 over 540) showed a period change larger than 5\%, the most relevant case being T UMi which is undergoing a large period and amplitude decrease. 

While the odds of finding another T UMi case were rather low, we decided to see if any of the Miras detected in the Asiago plates had a significant period or amplitude variation at 50 years distance. To this purpose, 
from February 2020 to April 2021 we performed a monitoring of 7 of these Mira stars in the R$_C$ and $i'_{Sloan}$ bands: the stars were selected on the basis that none of them was present in the ASAS-SN V-band public database.
Our star list is given in Table \ref{tab1}, together with their IPHAS2 catalog \citep{Barentsen14} J2000 coordinates (nearly identical to the 2MASS ones); the spectral type, period and average $i'_{Sloan}$ magnitude are 
taken from \citet{Nesci18}; the last two columns give the number of nights of our monitoring and the number of comparison stars used for our photometry. We remark that 3 stars are of spectral type S and 4 stars of spectral type M: none of 5 the carbon stars in the \citet{Nesci18} sample was a Mira variable.

\begin{table} 
\caption{Program stars}\vspace{3mm}  
\centering
\begin{tabular}{l c r r r r r}
\hline
Coordinates & SIMBAD & Spec. &Period & $i'_{Sloan}$ &   n.obs& comp \\
J2000           & name     &  type   & days   &   mag          &            &           \\
\hline
\hline
J004531.42 +595953.0 & [I81]M514            &S7e     & 377 & 11.52   & 37 & 52\\
J005302.81 +603547.6 &OT-Cas                &M4/5e  &292  & 11.07  & 40  & 40\\
J005933.99 +604318.4 &AV-Cas                 &M5/6e &330  &  10.43 & 35 & 33\\
J010642.86 +595819.2 &MIS-V1305           &M7      & 287  &  10.94 & 37 & 43\\
J010744.58 +590301.9 &V890-Cas            &S4/4    & 485  &13.15  & 41 &15  \\
J011225.70 +614146.3 &no-name               &M5/6   &181  &  12.61 & 38 & 22\\
J011259.79 +621046.8 &V418-Cas             &S7/2e  & 482 & 10.63  & 35 & 32\\ 	  
\hline
\end{tabular}\label{tab1}
\end{table}

After our project was started, the ZTF collaboration  \citep{Masci19} began to publish their data with free access via the IRSA website \citep{ZTF21}: an extensive database of variable stars (including ours)  based on the Data Release 2 (DR2) was published  in July 2020 by \citet{Chen20} with light curve from the year April 2018 to December 2019. The last ZTF release (DR5) covers our stars from  April 2018 up to December 2020, with a large overlap with our data allowing a consistency check, and to extend backwards the time interval by 1.6 years. V418 Cas however is poorly covered by ZTF-DR5, so our dataset for this star is particularly relevant.

In Section 2 we describe the instrumental setup used and the comparison stars sequences, in Section 3 the data analysis and results, in Section 4 our conclusions.
 
\section{Observations}

The observation were performed with 3 small telescopes located on the Subasio Mount near Assisi: a 150/650 mm Newtonian reflector, a 120/840 mm refractor, and a 70/420 mm refractor. The telescope turnovers were due to mechanical problems in their mountings: however, the observations were alway performed with the same filters and camera: a CCD Atic 16c-s, with a Sony ICX415AL sensor (782x582 square pixels of 8.3 micron), a Sloan $i'_{Sloan}$ and a Bessell $R_C$ filters. Exposure times ranged from 120 to 240 s. The scales of our images were therefore   2.63, 2.04, and 4.07 arcsec/pixel respectively. 
The logbook of the dates of use of the 3 telescopes and their focal plane scales in arcsec/pixel are reported in Table \ref{tab2}.

\begin{table} 
\caption{Logbook of the telescopes used.}\vspace{3mm} 
\centering
\begin{tabular}{ r c c c}
\hline
 telescope& date start & date end& scale ("/pix) \\
 \hline
 \hline
150/650 &2020-02-12 &2020-04-20& 2.63\\
 70/420 &2020-04-21 &2020-08-09& 4.07\\
120/840 &2020-08-09 &2020-11-19& 2.04\\
150/650 &2020-11-25 &2021-04-16& 2.63\\
\hline
\end{tabular}\label{tab2}
\end{table}

For each star 35 to 41 observations in each filter were secured, as given in Table \ref{tab1}. 

For each variable star, a large number of comparison stars within the field of view (see Table \ref{tab1}) were taken from the UCAC4 \citep{Zac12} catalog using the $r'_{Sloan}$ and the $i'_{Sloan}$ magnitudes. Therefore our Red magnitudes have a Sloan zero-point and will be referred in the following as $r'_{Sloan}$. We expect a small systematic offset with respect to the true $r'_{Sloan}$ magnitudes, due to the different passband of the $R_C$ and $r'_{Sloan}$ filters, and to the steep spectral slope of the Mira stars in the Red band: we will come back on this topic in Section 3.
For V890 Cas we selected the same comparison stars already used by \citet{Nesci16}.

All images were corrected for dark current and flatfield, and astrometrically calibrated with Astroart 7 \citep{astroart7}. Aperture photometry was performed wth IRAF-apphot, using an aperture about twice the FWHM of the stellar images (i.e. 2 to 3 pixels). The relation between instrumental and catalog magnitudes was always linear with slope very near to 1.00, as expected for a linear detector.
The stars were generally not detected if fainter than $r'_{Sloan}$ $\sim$16.0 mag, while were always well detected in the $i'_{Sloan}$ band. V890 Cas, the faintest of our stars, was detected in Red only near its maximum.

\begin{figure}
\centering
\includegraphics[width=10cm]{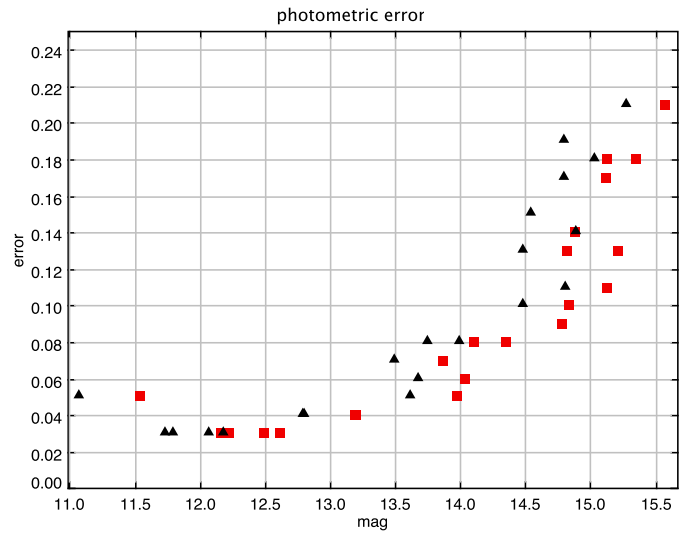} 
\caption{The photometric error as a function of the stellar magnitude: black points are $i'_{Sloan}$ data, red points are $r'_{Sloan}$ data.
}
\label{fig1}
\end{figure}

To evaluate the photometric uncertainty we computed, for each comparison star and for each image of a given field, its deviation $\epsilon_n$ from the linear fit of the catalog and instrumental magnitudes: the uncertainty $\sigma$ for each star was then computed as  $\sqrt{\sum (\epsilon_{n}^2/(N-1))}$ where N is the number of images. 
A plot of the error {\it vs} the star magnitude for one of our fields is given as example in Fig. \ref{fig1}: for bright magnitudes the error is essentially that of the input catalog, then it increases for fainter stars, as expected.  The trend is the same for both filters but the  $i'_{Sloan}$ error are a bit larger, likely due to the lower sensitivity of our camera. For each magnitude measure of the Mira variable  we adopted the uncertainty corresponding to its magnitude, as derived from the plot of its comparison stars.

The light curves of the 7 stars are shown in Fig. 2 to 8 together with the ZTF-DR5 data. It can be seen that
our $r'_{Sloan}$ magnitudes overlap very well the ZTF ones. Individual stars will be briefly discussed in Section 3.

\begin{figure}
\centering
\includegraphics[width=14cm]{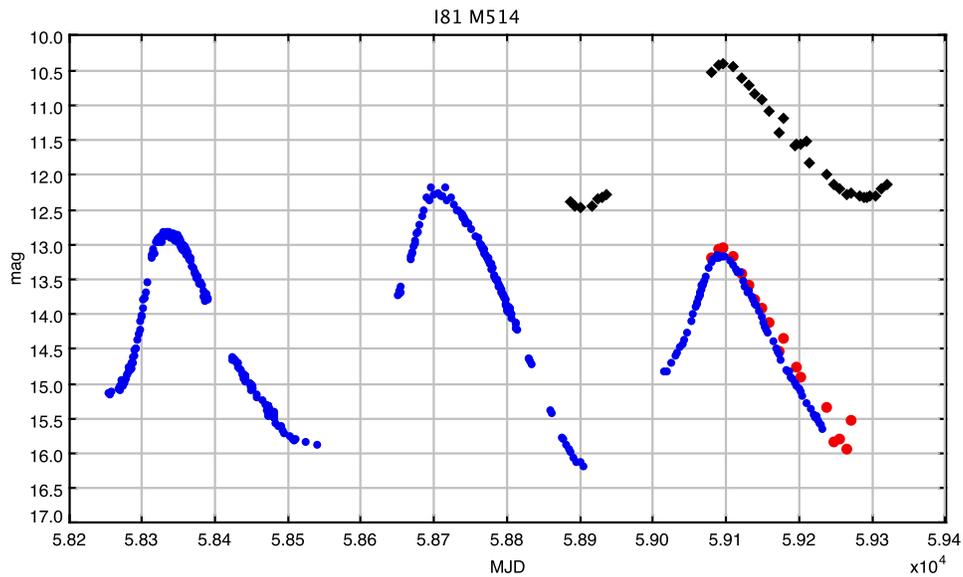} 
\caption{The light curve of [I81]M514: blue points are ZTF $r'_{Sloan}$ data, red points our $r'_{Sloan}$ data, black point our $i'_{Sloan}$.
}
\label{fig2}
\end{figure}

\begin{figure}
\centering
\includegraphics[width=14cm]{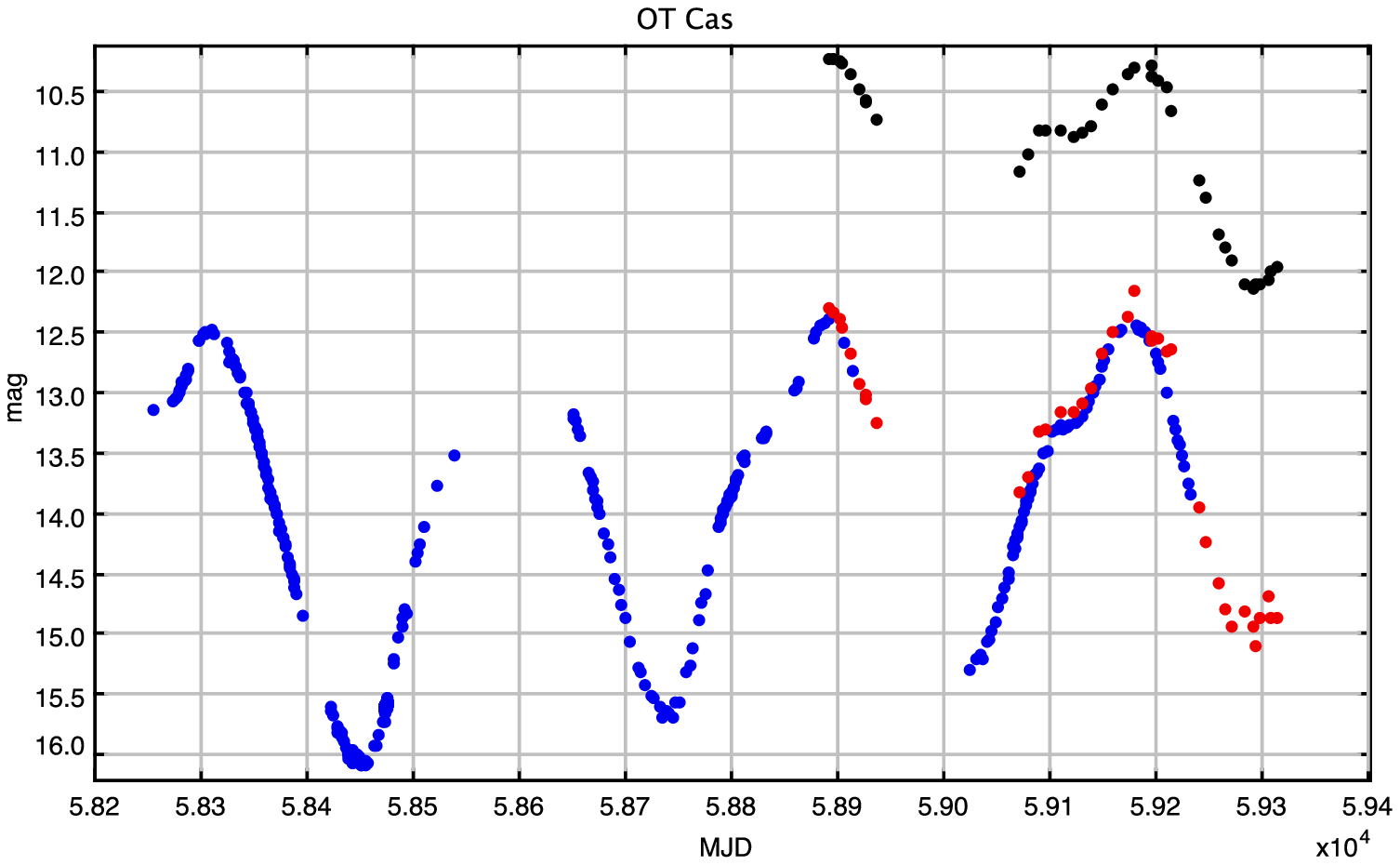} 
\caption{The light curve of OT Cas: blue points are ZTF $r'_{Sloan}$ data, 
red points our $r'_{Sloan}$ data, black point our $i'_{Sloan}$.
}
\label{fig3}
\end{figure}

\begin{figure}
\centering
\includegraphics[width=14cm]{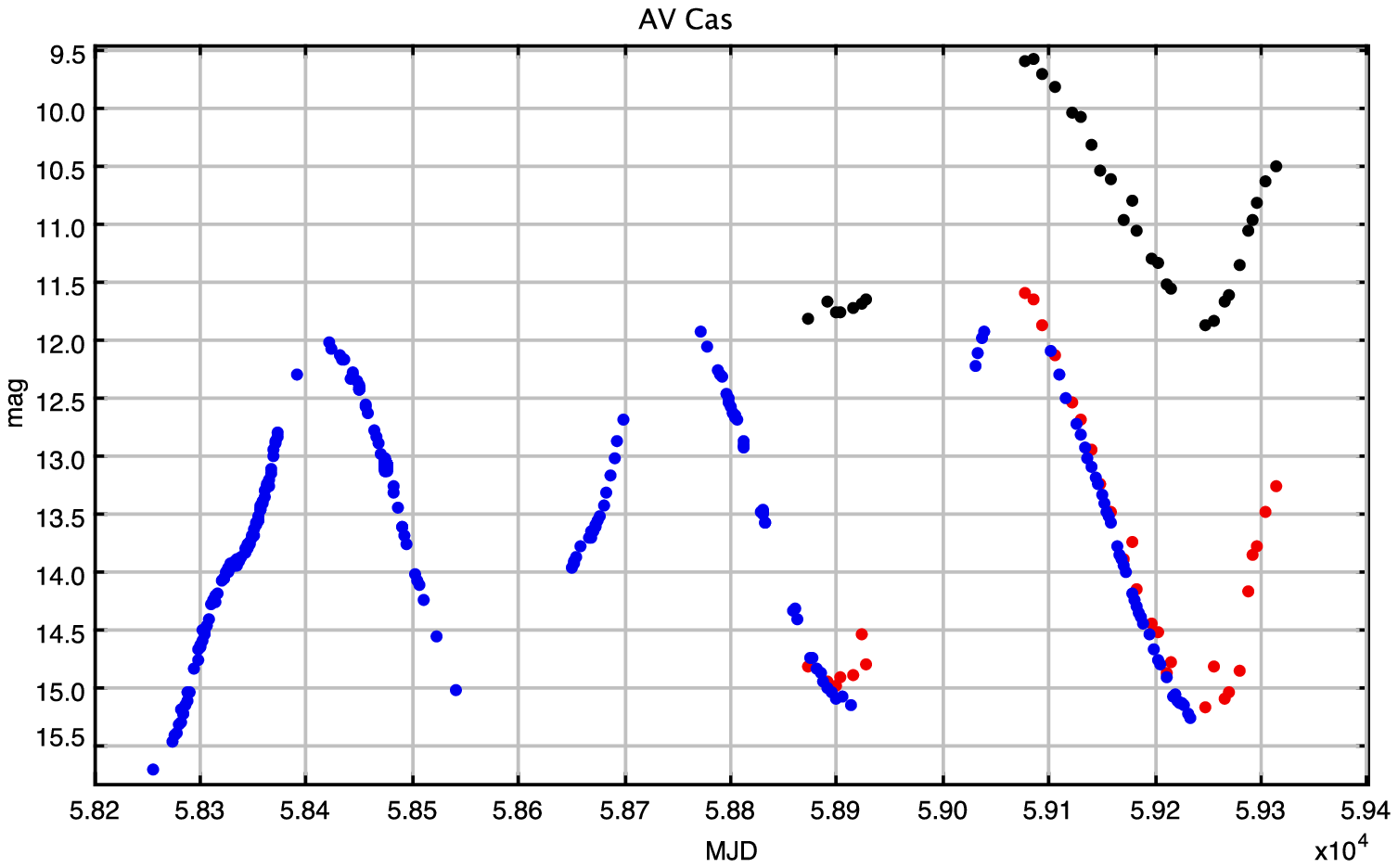} 
\caption{The light curve of AV Cas: blue points are ZTF $r'_{Sloan}$ data, 
red points our $r'_{Sloan}$ data, black point our $i'_{Sloan}$.
}
\label{fig4}
\end{figure}

\begin{figure}
\centering
\includegraphics[width=14cm]{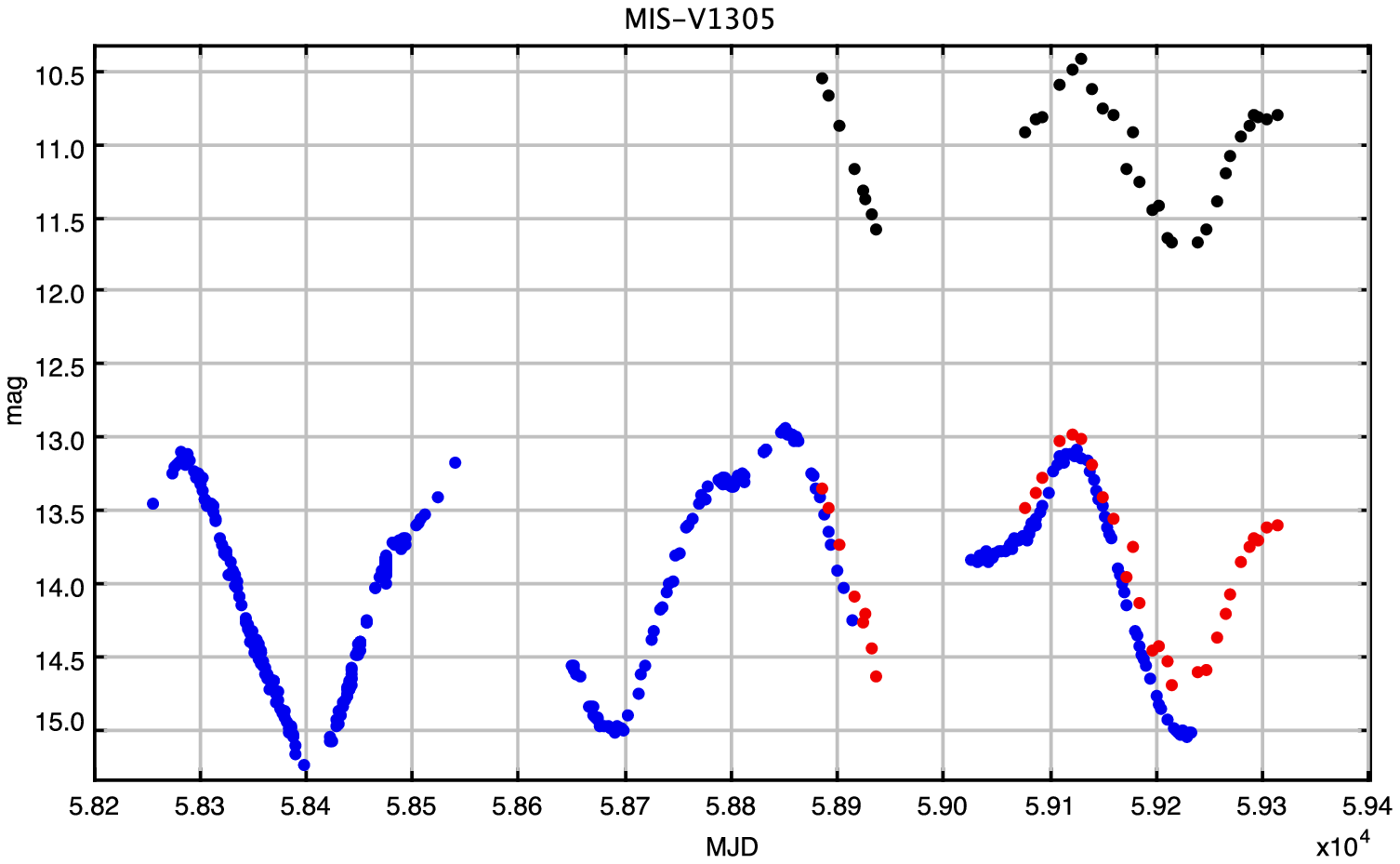} 
\caption{The light curve of MIS-V1305: blue points are ZTF $r'_{Sloan}$ data, red points our $r'_{Sloan}$ data, black point our $i'_{Sloan}$.
}
\label{fig5}
\end{figure}

\begin{figure}
\centering
\includegraphics[width=14cm]{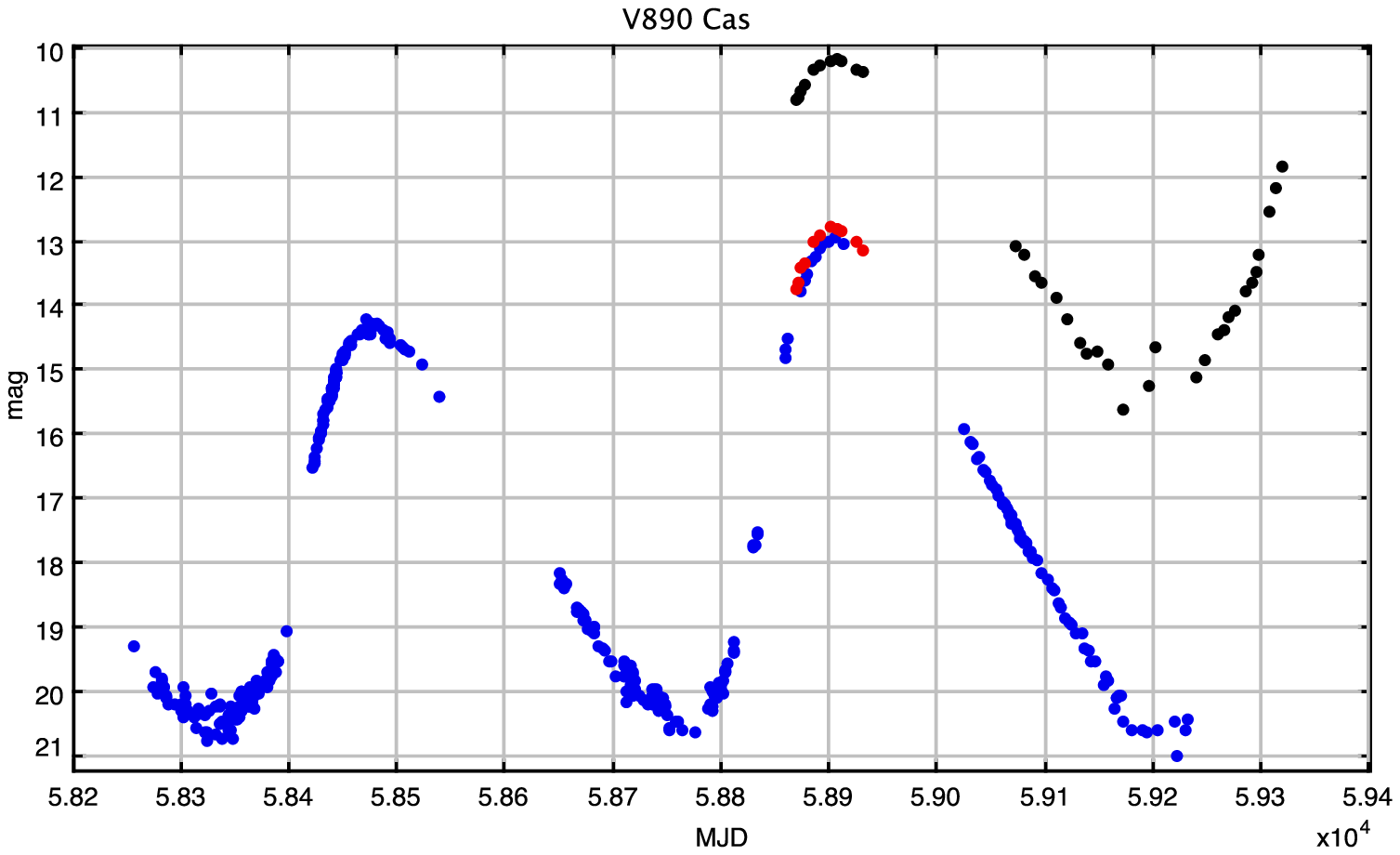} 
\caption{The light curve of V890 Cas: blue points are ZTF $r'_{Sloan}$ data, red points our $r'_{Sloan}$ data, black point our $i'_{Sloan}$.
}
\label{fig6}
\end{figure}

\begin{figure}
\centering
\includegraphics[width=14cm]{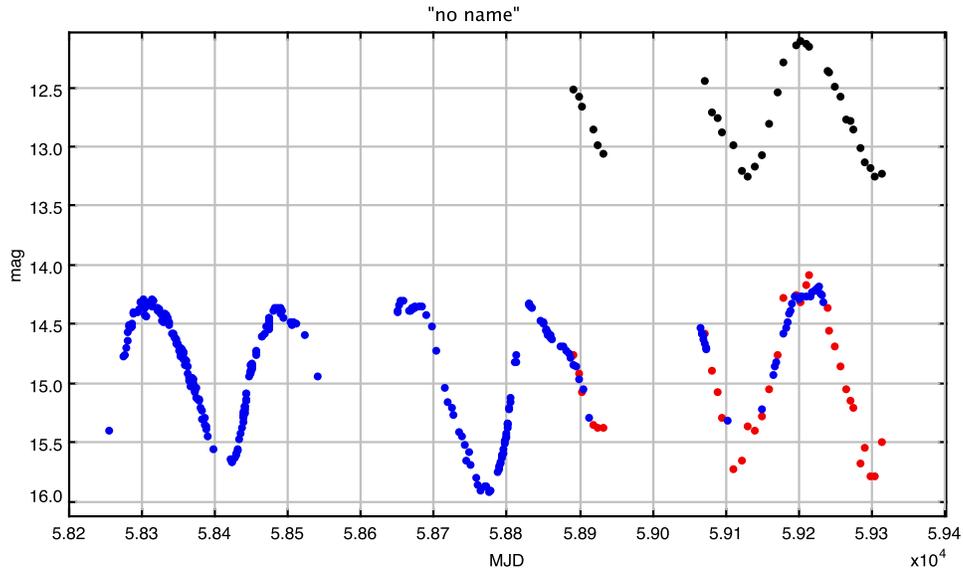} 
\caption{The light curve of 2MASS J011225.70+614146.3: blue points are ZTF $r'_{Sloan}$ data, red points our $r'_{Sloan}$ data, black point our $i'_{Sloan}$.
}
\label{fig7}
\end{figure}

\begin{figure}
\centering
\includegraphics[width=14cm]{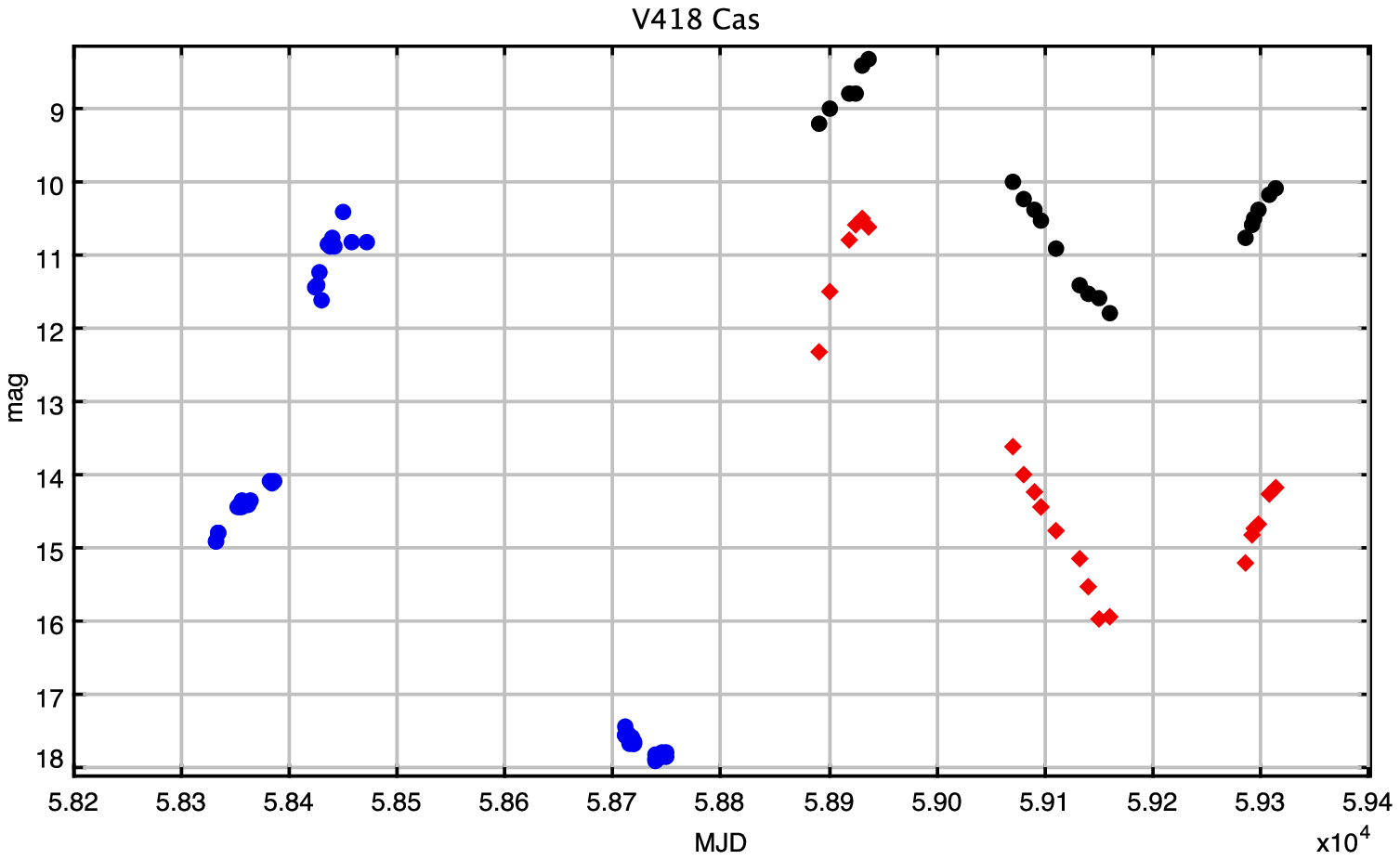} 
\caption{The light curve of V418 Cas: blue points are ZTF $r'_{Sloan}$ data, red points our $r'_{Sloan}$ data, black point our $i'_{Sloan}$.
}
\label{fig8}
\end{figure}

\section{Data analysis and Results}

\subsection{Periods}
 To derive the stars periods we used the $r'_{Sloan}$ magnitudes of the ZTF-DR2 database, save in the case of V418 Cas, which was poorly sampled in ZTF (see Fig. \ref{fig8}), so we merged our $r'_{Sloan}$ magnitudes to have a longer time coverage.
The periods were computed with the ANOVA code, based on the method by \citet{Schwarz96} as available in Peranso 2.60 \citep{Paunzen16}. For consistency of comparison, we used the same code also to recompute the periods of the old Asiago light-curves: the results were always fully consistent within a few days. The ZTF and the Asiago periods and epochs are reported in Table \ref{tab4}, as well as the Asiago periods computed with the Discrete Fourier Transform (DFT, \citet{Deeming75}) in the original paper by \citet{Nesci18}.
We recall that the photographic Asiago data cover a time interval of about 8 years (1967-1975), with a few sparse points in 1982-84, but with a photometric accuracy of about 0.15 mag, while the ZTF CCD data cover only 2.6 years but have a photometric accuracy about ten times better.

In the same Table \ref{tab4} we report the peak to peak amplitude of the $i'_{Sloan}$ light curves of the present and of the Asiago datasets.

\begin{table} 
\caption{Epochs, Periods, and amplitudes from the Asiago and the present datasets.}\vspace{3mm} 
\begin{tabular}{lccccccccc}
\hline
name & T0-Asiago & P-Asiago & P-Asiago&T0-now & P-now & P$_{peak}$&({\it d}lnP/{\it dt})$^{-1}$&Present& Asiago \\
          & MJD          &   ANOVA  &  DFT.     &  MJD    & days    &  days         &    yr       &ampl.  & ampl.    \\
\hline
\hline
$[I81]$M514 & 42120 &383(3) & 377 &59090 & 374(10) &377.1 &2094 & 2.1 & 3.3 \\
OTCas          & 42085 &294(4) & 292 &58310 & 283(4)   &295.0 &1341  &1.9 & 2.6  \\
AVCas          & 42090 &327(2) & 330 &59070 & 331(4)   &332.9 & 4162 &3.6 & 2.5 \\
MIS-V1305   & 42450 &287(1) & 287 &59120 & 279(3)   & 287.4 &1793 &1.3 & 2.0 \\
V890Cas      & 42150 &483(4) & 485 &58910 & 428(10)  &  ---    & 436 &5.5 &  5.6\\
no-name        & 42110 & 182(1)& 181 &58310 & 179(1)   &182.0 &3033  &1.1 & 1.4  \\
V418Cas      & 42640 & 477(5)& 482 &58450 & 473(6)   &479.1 & 5980 &4.3 & 5.6   \\
\hline
\end{tabular}
\label{tab4}
\end{table}

As a check of period variation, we report in column 7 a further period estimate P$_{max}$, computed assuming an integer number of periods between the Asiago and the recent epochs of maximum. For six stars this integer number resulted to be the same either using the Asiago or the new period determination, as expected if the periods were constant:  in the case of V890 Cas the number of cycles is very different for the two period measures (35 and 39), strenghtening the evidence of period variation.

For comparison with the work by \citet{Templeton05}, we computed for our stars the parameter {\it d}lnP/{\it dt}, defined as $[|(P_{Asiago} -P_{now})|]/[P_{mean} \times \Delta(T)$], where $\Delta$(T) is the time distance 
between the Asiago and the present datasets (50 years): the inverse of this quantity, in years, is listed in column 
8 of Table \ref{tab4}. The larger this number, the smaller the period variability.

From Table \ref{tab4} it is apparent a significant decrease in the period of V890 Cas (12\%), while for the other 6 stars the difference is below 3\%. Comparing the period variability rate (column 8) with the corresponding values in the \citet{Templeton05} sample, V890 Cas would rank in fourth position among the most variable Mira stars, between BH Cru and DF Her.

Regarding the light curve amplitudes, the present observations show for all stars, save AV Cas, an amplitude smaller than the historical ones; this could be due to the better accuracy of the modern CCD observations, but the amplitude of Mira variables is not strictly constant, so it could just be a consequence of the longer Asiago time coverage. 

Our very small sample shows that stars of longer periods have larger amplitudes, in agreement with the finding by Ita et al. (2021) from a much larger sample of Mira stars in the Small Magellanic Cloud.

\subsection{Luminosities and Color indexes}

As told in Section 2, our $r'_{Sloan}$ magnitudes may have a small systematic offset due to the actual use of an $R_C$ filter. To evaluate this offset we looked at the maxima of our light curves overlapped to the ZTF ones.
The differences for each star are reported in column 2 of Table \ref{tab3}: the average value is 0.10, quite small compared to the variation amplitude of the light curves.

For all the stars the color index ($r'_{Sloan}$-$i'_{Sloan}$) computed from our dataset showed a linear trend with the luminosity variation, of the form: 

\begin{equation}\label{eq1}
	(r'_{Sloan}-i'_{Sloan}) = a\times(r'_{Sloan}-rp_{Sloan})+ b,
\end{equation}

where ($r'_{Sloan}$-$rp_{Sloan}$) is the magnitude difference from the peak value. For V890 Cas, which was generally not detected by us in red, we computed the color index using the ZTF $r'_{Sloan}$ magnitudes nearly simultaneous to our $i'_{Sloan}$ observations. The slope (a) of the linear fit for each star is given in Table \ref{tab3}, as well as the color index at maximum (b), the correlation coefficient of the fit (always very good), the number of useful observations, and the spectral type reported from Table \ref{tab1}. 

We report also in the last columns of Table \ref{tab3}  the mean $i'_{Sloan}$ magnitudes in the Asiago and in our dataset. For 5 stars these magnitudes are fairly similar, indicating that no major changes occurred at 50 years distance. AV Cas and V418 Cas show a significant difference and will be discussed below.

\begin{table} 
\caption{Color trend with luminosity.}\vspace{3mm}
\centering 
\begin{tabular}{l r r r r r r c c}
\hline
SIMBAD & ZTF-our &a & b   & corr. &  n.obs & Spec& Asiago & present\\
 name     & r mag.   &    &mag& coef.&           &type  & mag.    & mag  \\
\hline
\hline
$[I81]$M514  &0.12 &0.328 & 2.66 & 0.98 & 19 &S7e    & 11.52&11.73\\
OTCas          &0.08 &0.324 & 2.03 & 0.94 & 38 & M4/5e &11.07&11.02\\
AVCas          & 0.05 &0.340 & 2.13 & 0.95 & 33 &M5/6e & 10.43& 11.00\\
MIS-V1305   & 0.12 &0.300 & 2.56 & 0.87 & 35 & M7     &10.94& 11.06 \\
V890 Cas      & 0.18 &0.346 & 2.62 & 0.98 & 34 & S4/4  &13.15&12.97 \\
no-name         & 0.08 &0.313 & 1.99 & 0.83 & 34 &M5/6  & 12.61 &12.75\\
V418Cas       &----- &0.453 & 2.10 & 0.96 & 21 & S7/2e  &10.63&10.20\\
\hline
\end{tabular}\label{tab3}
\end{table}

\subsection{Individual comments}

[I81]M514. This star still does not have a "standard" variable name of the form Vnnnn Cas. It is listed by \citet{Ichikawa81} as an S-type star, and it was tentatively identified in \citet{Nesci18} with the NSVS object \#1638444 in the on-line NSVS archive \citep{Wozniak04a}. This identification is incorrect. There are three possible counterparts of this star in the NSVS archive: \#1587497 at 0:45:32.5+59:59:52.98, with $V_{Rotse}$ declining from 11.5 to 12.9. A second object \#1638444 at 0:45:33.81+60:0:0.43, normally detected at $V_{Rotse}$=12.5 but with frequent drops down to r=15. A third object \#1665936 at 0:45:33.86+60:0:0.47, stable around $V_{Rotse}$=12.3 but with a single drop at 15.9. At 11.7 arcsec to 
the East of the Mira variable, inside the FWHM of the NSVS images  (20 arcsec), there is a non variable star (PanSTARRS r=14.7, i=14.5)  which is brighter than the Mira for nearly half of its period. This star was always well separated from the variable in our images, also those of the 70/420 telescope. This complex situation likely prevented the inclusion of this star in the final catalog of Red Variables of the NSVS survey     \citep{Wozniak04b}. The first source, \#1587497, has the nearest coordinates to the actual Mira position, and is the brightest, so it is the most appropriate cross-identification.
We looked also for a further period determination of this star using the maximum of the NSVS light curve: unfortunately, the limited time coverage (from MJD 51320 to 51620) and the presence of large holes in the sampling do not allow to safely establish the epoch of maximum.

OT Cas. The light curve shows a clear bump in the ascending branch, which is even more marked in the $i'_{Sloan}$ band. The minima show a monotonically rising trend while the maxima have a constant level. Opposite trends were present in the Asiago data, but the sampling was definitely lower (83 points in 8 years).

AV Cas. Its light curve is rather regular and the recent amplitude is larger than the historical one: this is rather peculiar because for all other stars we found that the present amplitude is smaller than the historical one. This fact may be linked to the fainter mean level of its recent light curve.

MIS-V1305.  There is a very nearby  (5 arcsec) star  at 01:06:43.41 +59:58:16.1 with magnitudes r=15.89 i=15.71 which is not resolved in our images. Our $i'_{Sloan}$  light curve is marginally affected by its 
presence around the minimum, while our $r'_{Sloan}$ is significantly flattened, as can be seen in Fig. \ref{fig5}. 
In the MISAO catalog, its magnitude range is reported as 10.8-12.4 (unfiltered); the MISAO catalog was made with telescopes of about 1 meter focal length and a KAF1600 CCD detector \citep{Yoshida99}, so also in their images the two stars were not resolved.
The star was indicated as a double period in \citet{Nesci18}, of 286 and 514 days, to explain the apparently alternate cycles of smaller and larger amplitude in the 8 years long Asiago lightcurve. 
The recent ZTF and our observations (Fig. \ref{fig5}) cover only 1100 days with four maxima and three minima, not enough to safely establish the presence of the 514 days period. However, the modern CCD light curve, which has a photometric accuracy much better than the old photographic one, shows a clear bump in the ascending branch, of variable shape: this bump may explain the fact that the Asiago light curve, when phased with the 287 days period, shows a large spread in the ascending branch, definitely larger than the descending one, and also the need of a second period in a DFT analysis of the Asiago light curve.

V890 Cas. The peak luminosity of the light curve is different in the two cycles covered by this monitoring. This was also shown by the Asiago light curve \citep{Nesci16}. The next maximum is expected in May 2021, when the star cannot be observed easily.

"no-name". This star is  identified in SIMBAD and in VSX with its 2MASS coordinates, and does not have still a "standard" variable name of the form 
Vnnnn Cas.

V418 Cas. The ZTF coverage is rather short, without overlap with our data, as shown in Fig. \ref{fig8}; only the sum of
the two datasets allowed to derive its present period, which is fully consistent  with the old one. The present mean $i'_{Sloan}$ magnitude is 0.4 mag brighter than the Asiago one, but is computed over less than one cycle.

\section{Conclusions}\label{secconc}
We derived the $r'_{Sloan}$  and $i'_{Sloan}$ light curves, between February 2020 and April 2021, of 7 Mira stars in a sky area centered on Gamma Cas, whose historical (1967-84) light curves were obtained from Asiago  infrared plates \citep{Nesci18}. One star (V890 Cas) was detected by us in the red band only around its maximum, due to its faintness.

The recent publication of the ZTF-DR5 data allowed to get the $r'_{Sloan}$ light curves of these stars between April 2018 and December 2020, with large overlap with our data, which were found largely consistent.

We computed for all the stars their period and epoch of maximum form the (ZTF+ours) $r'_{Sloan}$ data using the ANOVA code \citep{Paunzen16}.
The star  V890 Cas, one of our two stars with period longer than 450 days, showed a marked period shortening (12\%) putting this stars among those with the largest period change known. For the other stars, the recent periods were consistent with the old ones at a 3\% level or better.

For 5 stars, the mean $i'_{Sloan}$ magnitude was found consistent with the Asiago one, while the variation amplitude was generally smaller. Of the two stars with different mean magnitude, V418 Cas was observed in $i'_{Sloan}$ for just one period, while AV Cas is the only star with a present variation amplitude larger than the historical one.

The color index ($r'_{Sloan}$ - $i'_{Sloan}$) of all stars showed a marked correlation with the star magnitude, being bluer when brighter, with tipical variation of 0.32 mag for a 1.00 $r'_{Sloan}$ magnitude variation. This is in qualitative agreement with the expectations from previous works, like e.g. \citet{Celis77} in the optical $U,B,V$ bands for  Milky Way Miras, or \citet{Ita21} in the$ I,J,H,K_s$ bands for  Miras in the Small Magellanic Cloud.

\setcounter{secnumdepth}{0}
\OEJVacknowledgements{We thank dr. Corinne Rossi for useful suggestions. This paper made use of the ASASSN, CDS, IRSA, NSVS, VSX, and ZTF databases.
Partially based on observations obtained with the Samuel Oschin 48-inch Telescope at the Palomar Observatory as part of the Zwicky Transient Facility project. ZTF is supported by the National Science Foundation under Grant No. AST-1440341 and a collaboration including Caltech, IPAC, the Weizmann Institute for Science, the Oskar Klein Center at Stockholm University, the University of Maryland, the University of Washington, Deutsches Elektronen-Synchrotron and Humboldt University, Los Alamos National Laboratories, the TANGO Consortium of Taiwan, the University of Wisconsin at Milwaukee, and Lawrence Berkeley National Laboratories. Operations are conducted by COO, IPAC, and UW. 
} \\


\end{document}